\documentclass[10pt,conference]{IEEEtran}

\usepackage{enumitem}
\usepackage{footnote}
\usepackage{amsmath}
\usepackage{array}
\usepackage{graphicx}
\usepackage[dvipsnames]{xcolor}
\usepackage{lscape}
\usepackage{subcaption}
\usepackage{soul}
\usepackage{pgfplots}
\usepackage{amsbsy}
\usepackage{tikzit}
\usepackage{makecell}
\usepackage{blindtext}
\usepackage{booktabs}
\usepackage{tcolorbox}

\pgfplotsset{compat=1.16}

\usepackage[switch]{lineno}

\def\BibTeX{{\rm B\kern-.05em{\sc i\kern-.025em b}\kern-.08emT\kern-.1667em\lower.7ex\hbox{E}\kern-.125emX}}

\graphicspath{ {figures/} }
\def\tablename{Table}
\usepackage[labelfont=bf]{caption}

\begin{document}
\title{SoftNER: Mining Knowledge Graphs From\\Cloud Incidents}

\author{
\IEEEauthorblockN{Manish Shetty\IEEEauthorrefmark{1}, Chetan Bansal\IEEEauthorrefmark{1}, Sumit Kumar\IEEEauthorrefmark{2}, \\ Nikitha Rao\IEEEauthorrefmark{1}, Nachiappan Nagappan\IEEEauthorrefmark{1} 
}
\IEEEauthorblockA{\IEEEauthorrefmark{1}\textit{Microsoft Research}}
\IEEEauthorblockA{\IEEEauthorrefmark{2}\textit{Microsoft}}
}

\newcommand{\softner}{SoftNER}
\newcommand{\CompanyX}{Microsoft}
\newcommand{\todo}[1]{\textbf{\textcolor{red}{TODO: #1 }}}
\newcommand{\addRef}{\todo{REF HERE}}
\newcommand{\reduceVSpace}{\vspace{-4mm}}

\newcommand{\checkx}[1]{\textbf{\textcolor{orange}{CHECK: #1 }}}
\newcommand{\del}[1]{\textcolor{red}{\st{#1}}}
\newcommand{\add}[1]{\textcolor{blue}{#1}}

\maketitle

%
\begin{abstract}
The move from boxed products to services and the widespread adoption of cloud computing has had a huge impact on the software development life cycle and DevOps processes. Particularly, incident management has become critical for developing and operating large-scale services. Prior work on incident management has heavily focused on the challenges with incident triaging and de-duplication. In this work, we address the fundamental problem of structured knowledge extraction from service incidents. We have built \softner{}, a framework for mining Knowledge Graphs from incident reports. First, we build a novel multi-task learning based BiLSTM-CRF model which leverages not just the semantic context but also the data-types for extracting factual information in the form of named entities. Next, we present an approach to mine relations between the named entities for automatically constructing knowledge graphs. We have deployed \softner{} at \CompanyX{}, a major cloud service provider and have evaluated it on more than 2 months of cloud incidents. We show that the unsupervised machine learning pipeline has a high precision of 0.96. Our multi-task learning based deep learning model also outperforms the state-of-the-art NER models. Lastly, using the knowledge extracted by \softner{}, we are able to build accurate models for applications such as incident triaging and recommending entities based on their relevance to incident titles.
\end{abstract}

\section{Introduction}
\label{sec:introduction}

In the last decade, two major paradigm shifts have revolutionized the software industry. First is the move from boxed software products to \emph{services}. Large software organizations like Adobe and Microsoft\footnote{https://www.pcworld.com/article/2038194/microsoft-says-its-boxed-software-probably-will-be-gone-within-a-decade.html} which pre-date the internet revolution have been aggressively trying to move from selling boxed products to subscription based services. This has primarily been driven by the benefits of subscription based services such as scalability, faster releases, insights from telemetry, and, better revenue stability. The second major shift has been the widespread adoption of \emph{public clouds}. More and more software companies are moving from on-premises data centers to public clouds like Amazon AWS, Google Cloud, Microsoft Azure, etc. Gartner has forecasted\footnote{https://www.gartner.com/en/newsroom/press-releases/2019-11-13-gartner-forecasts-worldwide-public-cloud-revenue-to-grow-17-percent-in-2020} the public cloud market to grow to about \$$266$ billion in revenue in $2020$, out of which about 43\% revenue will be from the Software as a Service (SaaS) segment. The cloud revolution has enabled companies like Netflix, Uber, etc. to build internet scale products without having to provision their own infrastructure.

These paradigm shifts have also had a transformational effect on the way software is developed, deployed, and maintained. For instance, software engineers no longer develop monolithic software. They build services that have dependencies on several $1^{st}$ party and $3^{rd}$ party services and APIs. Typically, any web application will leverage cloud services for basic building blocks like storage (relational and blob), compute, and authentication. These complex dependencies introduce a bottleneck where a single failure can have a cascading effect. In 2017, a small typo led to a major outage in the AWS S3 storage service\footnote{https://www.wsj.com/articles/amazon-finds-the-cause-of-its-aws-outage-a-typo-1488490506}, which ended up costing over \$$150$ million to customers like Slack, Medium, etc. Microsoft had a glitch in their Active Directory in October 2019\footnote{https://www.theregister.co.uk/2019/10/18/microsoft\_azure\_mfa/}, which locked out customers from accessing their Office 365 and Azure accounts. These outages are inevitable and can be caused by various factors such as code bugs, misconfigurations \cite{mehta2020rex}, or even environmental factors.

To keep up with these changes, DevOps processes and platforms have also evolved over time \cite{dang2019aiops}, \cite{kumar2019building}. Most software companies these days have incident management and on-call duty as a part of the DevOps workflow, where the key motivation is to reduce impact on customers by mitigating any issue as soon as possible. We discuss the incident life-cycle and some of the associated challenges in detail in Section~2. Prior work on incident management has largely focused on two challenges: incident triaging \cite{ContinuousTriageASE2019}, \cite{EmpiricalIcMICSE2019} and diagnosis \cite{bansal2019decaf}, \cite{luo2014correlating}. Here, triaging refers to the process of identifying and routing the incident to the appropriate team for resolution. Chen et al. \cite{ContinuousTriageASE2019} did an empirical study where they found that upto 60\% of incidents can be mis-triaged. They proposed DeepCT, a deep learning approach for automated incident triaging using incident data (title, summary, comments) and environmental factors.

Based on our experience from operating web-scale cloud services at \CompanyX{} and discussions with various product teams, we observe that a key challenge in incident management lies in the lack of structured representations of these incidents. These incidents could be created by a wide variety of sources such as customers, engineers, and even automated monitoring systems. They are mostly unstructured and contain several types of information like incident metadata, description, stack traces, outputs of shell scripts, images, etc. As a result, on-call engineers spend a considerable amount of effort manually parsing verbose incident descriptions to understand the issue, locating key information for mitigation, and finally engaging the appropriate team to acknowledge and fix the issue. The time and effort spent here is termed as Time-To-Engage (TTE) and adds a significant delay to the ensuing tasks in the incident life-cycle as described in Section \ref{sec:incident-lifecycle}.

Thus, in this work, we address the key problem of \textbf{extracting structured knowledge from service incidents}. This structured knowledge would reduce the effort spent by on-call engineers by opening up avenues for automating processes like log extraction and health checks on resources (VMs, Databases, etc.) identified within these descriptions. The extracted knowledge would also help build better models for performing downstream tasks like triaging and root-cause analysis. Please refer to Section \ref{sec:applications} and Section \ref{sec:discussion} to further understand the applications of our framework.

Ideally, any knowledge extraction framework should have the following qualities:
\begin{enumerate}
    \item It should be \textbf{unsupervised} because it is laborious and expensive to annotate a large amount of training data. This is important since a service's unique vocabulary is unknown.
    \item It should be \textbf{domain agnostic} so that it can scale to a high volume and a large number of information entity types. Unlike in the web domain, where there is a small set of key entities such as people, places, and organizations, for incidents, we don't know these entities apriori.
    \item It should be \textbf{extensible} so that we can adapt the bootstrapping techniques to incidents from other services, or even other data sets such as bug reports. This is critical because each service (e.g. compute, networking, storage) could have its unique vocabulary and terminology.
\end{enumerate}

\begin{figure*}
\includegraphics[width=1\textwidth]{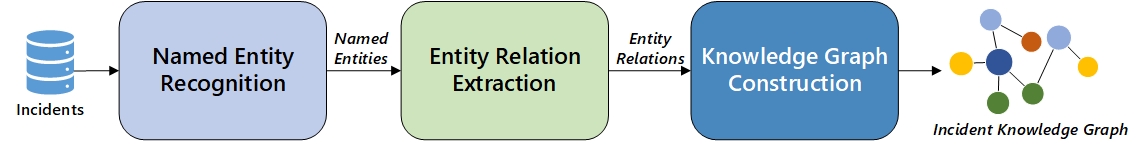}
\caption{\textbf{SoftNER Overview}}
\label{fig:overview}
\centering
\end{figure*}

We have designed \textbf{Soft}ware artifact K\textbf{N}owledge \textbf{E}xt\textbf{R}action (\softner{}), a framework for unsupervised knowledge extraction from service incidents, which has these three qualities: unsupervised, domain agnostic, and extensible. As shown in Figure \ref{fig:overview}, we break down knowledge extraction into three steps. First, we use Named-entity recognition (NER) for extraction of factual and structured information from the incidents. We leverage syntactic pattern extractors for bootstrapping the training data. Further, we incorporate a novel multi-task BiLSTM-CRF deep learning model with an attention mechanism. Next, we enrich these entities by mining binary relations between the entities. Lastly, we automatically construct knowledge graphs using the entities and relations extracted. We have evaluated and deployed \softner{} at \CompanyX{}, a major cloud service provider. We show that our unsupervised knowledge graph mining has high precision. Our multi-task deep learning model also outperforms existing state-of-the-art models on the NER task. Lastly, using the extracted knowledge, we show that we can build more accurate models for key downstream tasks like incident triaging and also improve tooling in incident management platforms. \\

In this work, we make the following main contributions:
\begin{enumerate}
    \item We propose \softner{}, the first approach for completely unsupervised Knowledge Graph mining from service incidents.
    \item We build a novel multi-task learning based deep learning model for named-entity recognition which leverages not just the semantic features but also the data-types. Our evaluation shows that it outperforms the existing state-of-the-art NER models. 
    \item We propose a novel approach to mine relations between the extracted entities for construction of knowledge graphs.
    \item We do an extensive evaluation of \softner{} on over 2 months of cloud service incidents from \CompanyX{}.\footnote{We cannot disclose the number of incidents due to Microsoft Policy.}
    \item Lastly, we have deployed \softner{} in production at \CompanyX{} where it has been used for knowledge extraction from incidents for over 6 months.
\end{enumerate}

The rest of the paper is organized as follows: In Section \ref{sec:incident-lifecycle}, we discuss insights from the incident management processes at \CompanyX{}. Section \ref{sec:overview} provides an overview of the \softner{} framework. Section \ref{sec:named-entity-recognition} and \ref{sec:rel-extraction-knowledge-graphs} provide details of our approach to knowledge extraction using named-entity recognition and knowledge graph construction respectively. Section \ref{sec:implementation} describes the implementation  and  deployment details. In Section \ref{sec:evaluation}, we discuss the experimental evaluation of our approach. Section \ref{sec:applications} describes two applications of \softner{} in detail and we discuss the generalizability and future work in Section \ref{sec:discussion}. Section \ref{sec:related-work} discusses the related work and we conclude the paper in Section \ref{sec:conclusion}.

This paper extends our prior publication~\cite{9402085} presented at the $43^{rd}$ International Conference on Software Engineering: Software Engineering in Practice. New materials with respect to the conference version include:

\begin{itemize}
\item We have extended the SoftNER framework by proposing and evaluating an unsupervised approach to further extract entity relations and constructing knowledge graphs using mutual information and co-occurrences (Section \ref{sec:rel-extraction-knowledge-graphs}).

\item We showcase an application of the constructed knowledge graph by recommending required entities for incidents based on incident titles (Section \ref{subsec:entity-incident-relevance}). Here, we use a combination of clustering and a novel path based scoring technique to identify entity-incident relevance.x

\item Additional details regarding the integration of the \softner{} framework with the incident management platform at \CompanyX{} are provided (Section \ref{sec:implementation}).
\end{itemize}
\section{Incident life-cycle}
\label{sec:incident-lifecycle}

\begin{figure}[h]
\includegraphics[width=1\linewidth]{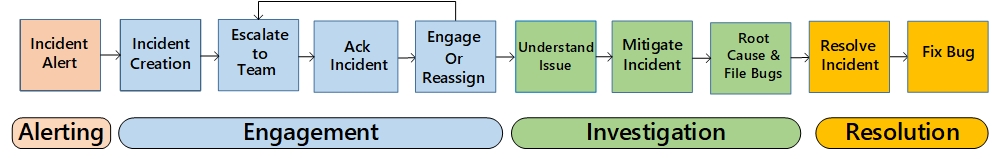}
\caption{\textbf{Incident Life-Cycle}}
\label{fig:incident-lifecycle}
\end{figure}

In \CompanyX{}, an incident is defined as an unplanned interruption or degradation of a product or service that is causing customer impact. For example, a slow connection, a timeout, a crash, etc. could constitute an incident. Here, we detail the incident management process which defines the various steps an incident goes through - from creation to closing. Bugs and incidents are very different, as in our case incidents may or may not lead to bugs. Furthermore, incidents often require the involvement of on-call developers who have been designated to respond to incidents. Figure \ref{fig:incident-lifecycle} presents a high level view of the incident management process. Most online services have their own specific incident management protocol. This figure is a generic process that could apply outside of \CompanyX{} as well.

The incident management process is broadly classified into four phases. In the first phase - \emph{alerting} phase, typically an alert is fired when the service monitoring metrics fall below a pre-defined acceptance level in terms of performance (e.g. slow response, slow transfer rate, system hang or crash, etc.). This leads to phase two - the \emph{engagement} phase. In this phase, an incident is first created in the incident database. It is then escalated to a ``related" team. The identification of the first ``related" team is automatic, based on heuristics or component ownership. The team investigates the incidents and engages with relevant stakeholders or re-routes it to a more appropriate team to repeat the steps. This process is called \textit{Incident Triage}, where the appropriate team is identified to take up the resolution of this incident. Following triage, in the \emph{investigation} phase, the appropriate team identifies the problem cause and moves over to mitigation and root cause analysis. Then, identified bugs, if any, are filed for engineering teams to fix. In the final phase of \emph{resolution}, the incident is resolved and bugs are fixed in the system. Our work applies directly to the engagement and investigation phases, dealing with unsupervised structured knowledge extraction from service incident descriptions.

\section{Overview}
\label{sec:overview}

\begin{figure*}
\includegraphics[width=1\textwidth]{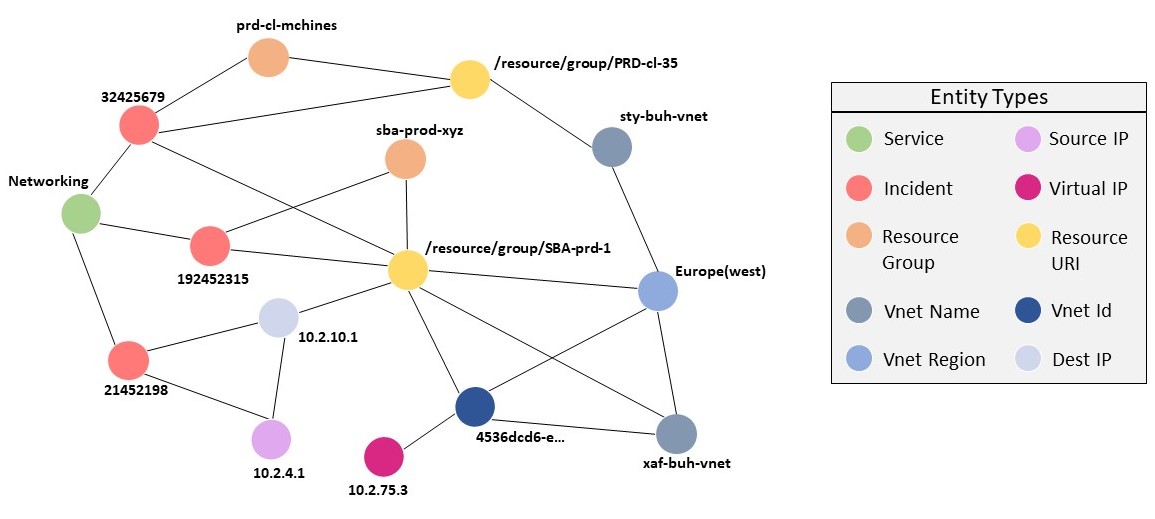}
\caption{\textbf{Incident Knowledge Graph}}
\label{fig:kgraph-overview}
\vspace{-1em}
\end{figure*}

Incident management is key to running large scale services in an efficient way. However, there is a lot of scope for optimization which can inturn increase customer satisfaction, reduce on-call fatigue, and provide revenue savings. Existing work on incident management has primarily focused on incident triaging. The state-of-the-art incident triaging methods \cite{ContinuousTriageASE2019} use novel deep learning methods which take raw unstructured incident descriptions as input. In this work, we focus on the fundamental problem of structured knowledge extraction from these unstructured incidents. With the structured information, we can save time and effort for on-call engineers by automating manual processes such as running automated health checks on identified resources as described in an example at the end of this section. At the same time, with this structured representation, we can build simpler yet better machine learning models for tasks like incident triaging.

To solve the challenges with incident management, we have designed the \softner{} framework. It is the first automated approach for structured knowledge extraction from service incidents. We frame the initial structured knowledge extraction problem as a \textit{Named-Entity Recognition} (NER) task, which has been well explored in the Information Retrieval (IR) domain \cite{nadeau2007survey}, \cite{lample2016neural}. Named-Entity Recognition is defined as the task of parsing unstructured text to not only detect entities but also classify them into specific categories. An entity can be any chunk of text which belongs to a given type or category. As an example, here is the input and output of a NER task for a news headline:

\medskip
\textit{\textbf{Input}}: Over 320 million people have visited the Disneyland in Paris since it opened in 1992.

\smallskip
\textit{\textbf{Output}}: Over \textbf{[320 million \textsubscript{COUNT}]} people have visited the \textbf{[Disneyland \textsubscript{ORG}]} in \textbf{[Paris \textsubscript{LOC}]} since it opened in \textbf{[1992 \textsubscript{YEAR}]}.

\medskip

Framing the knowledge extraction problem as a NER task enables us to not only extract factual information from the incidents but also classify them as specific entities. For instance, if we just extract a \textit{GUID} from the text, it provides limited context. However, identifying that \textit{GUID} as a \textit{Resource Id} is much more useful to on-call engineers, who can then identify affected resources, or to other models that perform tasks like triaging. Thus, making it more suitable for the incident management scenario when compared to other solutions like text summarization. 

One key limitation of any supervised machine learning pipeline is the requirement of huge amounts of labeled data which can be cost prohibitive to manually generate. In service incidents, the lack of existing training data prevents us from using any supervised or semi-supervised techniques. \softner{} uses pattern extractors which leverage the \emph{key-value} and \emph{tabular} structural patterns in the incident descriptions to bootstrap the training data. We then use label propagation to generalize the training data beyond these patterns. We also incorporate a novel multi-Task deep learning model that is able to extract named-entities from the unstructured incident descriptions with very high accuracy. The model not only leverages the semantic features but also the data type of the individual tokens (such as GUID, URI, Boolean, Numerical, IP Address, etc.).

\smallskip
\textbf{Example}: Let's consider a real incident reported by a customer of the Cloud Networking service operated by \CompanyX{}. The incident was caused due to an error in deleting a Virtual Network resource. The key information required to triage and mitigate this incident is actually the `\textit{Problem Type}' and the `\textit{Virtual Network Id}'. This information is already present in an unstructured form within the incident description and the challenge is to extract it automatically in a structured format. Using \softner{}, we can not only provide this key information to the on-call engineers but also automate some of the manual tasks such as running health checks. For instance, in this example, we can automatically look up the current status/logs of the resource using the identifier (\textit{Virtual Network Id}) before the on-call engineer engages.
\smallskip

Service incidents can be created by external customers or even automated monitoring systems. They contain unstructured information in various forms, like statements, conversations, stack traces, etc. As stated before, this makes incident descriptions rich in information that are identifiable as entities. Although extracting all entities is useful, certain entities are more important for the investigation and mitigation of an incident. To capture complete knowledge that can be used for other aspects, such as entity relevance, it is important to mine interactions and relations between entities. For instance, it would be quite useful to map a given \textit{Identifier} and \textit{IP Address} to the same \textit{Virtual Machine} resource. Another example, would be to identify \textit{Source} and \textit{Destination} entity relations in an incident dealing with physical networking devices. With these insights, \softner{} uses an unsupervised approach to extract entity relations using co-occurring entity pairs and pointwise mutual information. Having identified related entities, \softner{} automatically constructs an undirected incident knowledge graph. As shown in Figure \ref{fig:kgraph-overview}, the knowledge graph nodes represent cloud services, incidents, and entities extracted from incidents, and edges represent relatedness. The knowledge graph captures information that can be queried like traditional databases, analyzed like graph data structures, and allows inference of new knowledge.

Lastly, the structured knowledge extracted by \softner{} opens a wide range of applications. For instance, the extracted entities can be used as features to build more accurate learning models for predictive tasks like triaging, root causing, etc. The knowledge graph can also be utilized to improve tooling by intelligently recommending expected entities based on their relevance to the issue described in the incident. Below we list some of the terms which we will be using throughout the rest of the paper. These terms identify different aspects of the named-entities. 

\begin{itemize}
\item \textbf{Entity Name:} N-gram indicating the name of the entity. In the current implementation, N can range from 1 to 3.
\item \textbf{Entity Value:} The value of the named-entity for a given instance. 
\item \textbf{Data Type:} The data type of the values for the named-entity.
\item \textbf{Entity Relation:} A relationship between 2 or more entities.
\end{itemize}
\begin{figure*}
\includegraphics[width=1\textwidth]{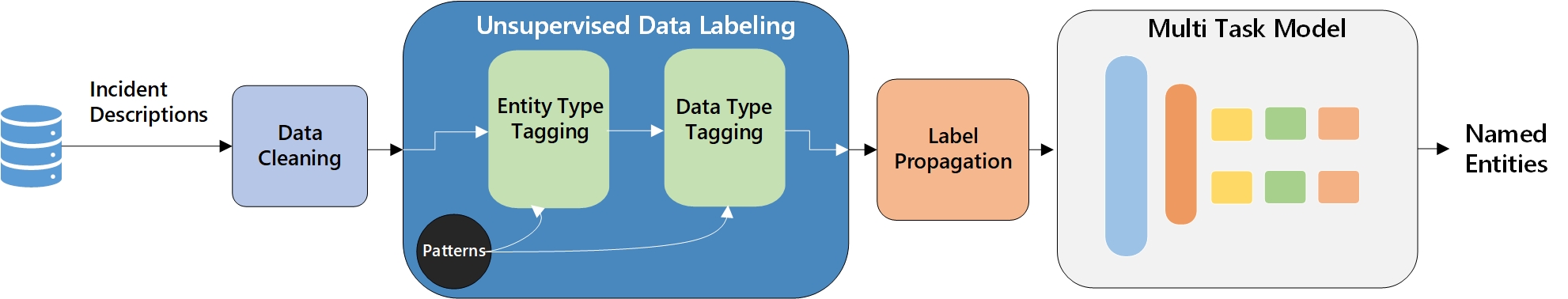}
\caption{\textbf{Named-Entity Recognition pipeline}}
\label{fig:pipeline}
\end{figure*}

\begin{table*}[!ht]
\centering
\caption{\textbf{Examples of entities extracted by \softner{}}}
\label{tab:entity-examples}
\begin{tabular}{lll}
\hline
\textbf{Entity Name} & \textbf{Data Type} & \textbf{Example} \\
\hline
Problem Type & Alphabetical & VNet Failure \\[-1pt]
Exception Message & Alphabetical & The vpn gateway deployment operation failed due to an intermittent error \\[-1pt]
Failed Operation Name & Alphabetical & Create and Mount Volume \\[-1pt]
Resource Id & URI & /resource/2aa3abc0-7986-1abc-a98b-443fd7245e6f/resourcegroups/cs-net/providers/network/frontdoor/ \\[-1pt]
Tenant Id & GUID & 4536dcd6-e2e1-3465-a22b-d25f62456233 \\[-1pt]
Vnet Id & GUID & 45ea1234-123b-7969-adaf-e0255045569e \\[-1pt]
Link With Details & URI & https://cloudx.com/caseview?cid=12\\[-1pt]
Source IP & IP Address & 198.168.0.1 \\[-1pt]
Status Code & Numeric & 500 \\[-1pt]
\hline
\end{tabular}
\end{table*}

\section{Named Entity Recognition}
\label{sec:named-entity-recognition}

Here, we describe our approach in implementing \softner{}'s named-entity recognition pipeline in detail. As shown in Figure \ref{fig:pipeline}, we start with the data cleaning process, followed by unsupervised data labeling. Then we describe the label propagation process and the architecture of the deep learning model.

\subsection{\textbf{Data Cleaning}}
Service incident descriptions and summaries are created by various sources such as external customers, feature engineers and even automated monitoring systems. The information could be in various forms, like textual statements, conversations, stack traces, shell scripts, images,  etc., all of which make the information unstructured and hard to interpret. Subsequently, we first prune tables in the incident description that have more than two columns and get rid of HTML tags using regexes and parsers. In this process, we also segment the information into sentences using newline characters. Next, we process individual sentences by cleaning up extra spaces and tokenize them into words. Our tokenization technique is custom implemented to handle camel-case tokens and URLs as well.

\subsection{\textbf{Unsupervised Data Labelling}}
\label{subsec:unsupervised_data_labelling}

A major challenge in our case, was the lack of a pre-existing labelled data set which can be used for a supervised NER task. It would also be very expensive to manually label data since entity types are unknown and also vary across different services. Thus, \softner{} uses an unsupervised framework to create a labeled corpus. Here are the steps followed to automatically generate a labeled corpus for named-entity extraction:

\smallskip
\subsubsection{Entity Type Tagging}
As mentioned above, we neither have a pre-existing labeled data set nor a predefined fixed set of entity types. Thus, in this phase, we first identify a candidate entity set. We then clean that set and eliminate noisy entities. This final set of entities is used to tag the incident data set.\\

\textbf{Step 1A (Candidate Identification)}: Since we don't have a list of entity types apriori, we first bootstrap the framework with a candidate set of entity name and value pairs that are identified without manual effort. For this, we have built pattern extractors using some structural patterns commonly found in the incident descriptions: 
\begin{itemize}
    \item \textbf{Key-Value pairs} - This pattern is commonly used in the incident descriptions to specify various entities where the Entity Type (key) and Value are joined by a separator such as `:'. For instance, ``\textit{Status code: 401}" or ``\textit{Problem type: VM not found}". Here, we split the sentence on the separator and extract the first half as the \textit{Entity Type} and the second half as the \textit{Entity Value}.
    \item \textbf{Tables} - Tables also occur quite frequently in incident descriptions, especially the ones which are created by bots or monitoring services.
    In a two-column html table, we extract the first column values as \textit{Entity Types} and the second as \textit{Entity Values}.
\end{itemize}

The above commonly occurring patterns are used to extract a candidate set of entity-value pairs after parsing the data set. Note that the above listed patterns do not constitute an exhaustive set. More patterns can be used to generate labeled data using methods like weak supervision \cite{ratner2017snorkel}, \cite{rao2020code}.\\

\textbf{Step 1B (Candidate Elimination)}: Now, we have a candidate set of entity names and values. However, the candidate set is noisy since we have extracted all text which satisfies these patterns. Thus, we filter out candidate entity names that contain symbols or numbers, as they are noisy labels. We further extract n-grams (n: 1 to 3) from the candidate entity names and take the top K most frequently occurring n-grams. Here, K is a parameter that can be chosen by the user to retrieve a fixed number of entity types. In this process, less frequent and thus noisy candidate entity types, such as \textit{``token acquisition started"} and \textit{``database connected"}, are pruned. We manually analyze the effect of K, in Section \ref{subsec:entity_type_evaluation}, using a precision-vs-rank plot.
Furthermore, with this n-gram approach, entity-value variations such as [\textit{``My Subscription Id", ``6572"}] and [\textit{``The Subscription Id", ``6572"}] would be transformed to [\textit{``Subscription Id", ``6572"}] since \textit{``Subscription Id"} is a commonly occurring bi-gram in the candidate set.

After performing the above steps, a final set of entity types, represented as n-grams, is determined. With this set, every occurrence of these n-grams, in the context of the 2 chosen patterns - key-value pairs and tables - are tagged in a single pass over the data set. Please refer to Table \ref{tab:entity-examples} for examples of entities extracted using the unsupervised approach.

\smallskip
\subsubsection{Data-Type Tagging}
For the refined candidate set, we next infer the data type of the entity values using in-built Python functions such as ``isnumeric" along with custom regexes. This step, in addition to the entity-type tagging, is leveraged in \softner{}'s multi-task learning model, where we jointly train to predict both the entity type and the data type. These tasks are complementary and help improve the accuracy for each of the individual prediction tasks. Based on discussions with the service engineers, we have defined the following data types:
\begin{itemize}
    \item \textbf{Basic Types}: Numeric, Boolean, Alphabetical, Alphanumeric, Non-Alphanumeric
    \item \textbf{Complex Types}: GUID, URI, IP Address
    \item \textbf{Other}
\end{itemize}
To infer the data type for a given entity, we compute it for each occurrence of a named entity in the data set. Then, conflicts are resolved by taking the most frequent type. For instance, if ``VM IP" entity is most commonly specified as an IP Address but sometimes is specified as a boolean, due to noise or dummy values, we correctly infer its data type as an IP Address.

\subsection{\textbf{Label Propagation}}
With the unsupervised tagging, we have bootstrapped the training data using pattern extractors and heuristics. While this allows us to generate a seed data set, the recall would suffer since the entities could occur outside the context of the chosen key-value or tabular patterns. In the absence of ground truth or labeled data, it's a nontrivial problem to solve. Thus, to avoid overfitting the deep learning model on specific patterns that were used to bootstrap labeled data, we would want to generalize or diversify the labels.

We use the process of \textit{label propagation} to solve this challenge. We use the entity values extracted in the bootstrapping process and propagate their types to the entire corpus. For instance, if the IP Address ``127.0.0.1" was extracted as a ``Source IP" entity, we would tag all untagged occurrences of ``127.0.0.1" in the corpus as ``Source IP". As we can imagine, there are certain edge cases that need to be handled. For instance, we cannot use this technique for entities with Boolean data type. It would also not work for entities whose values are descriptive. Lastly, it's possible that different occurrences of a particular value were tagged as different entities during bootstrapping. We resolve conflicts during \textit{label propagation} based on popularity, i.e., the value is tagged with the entity type which occurs more frequently across the corpus.


\begin{figure}[h]
\includegraphics[width=1\linewidth]{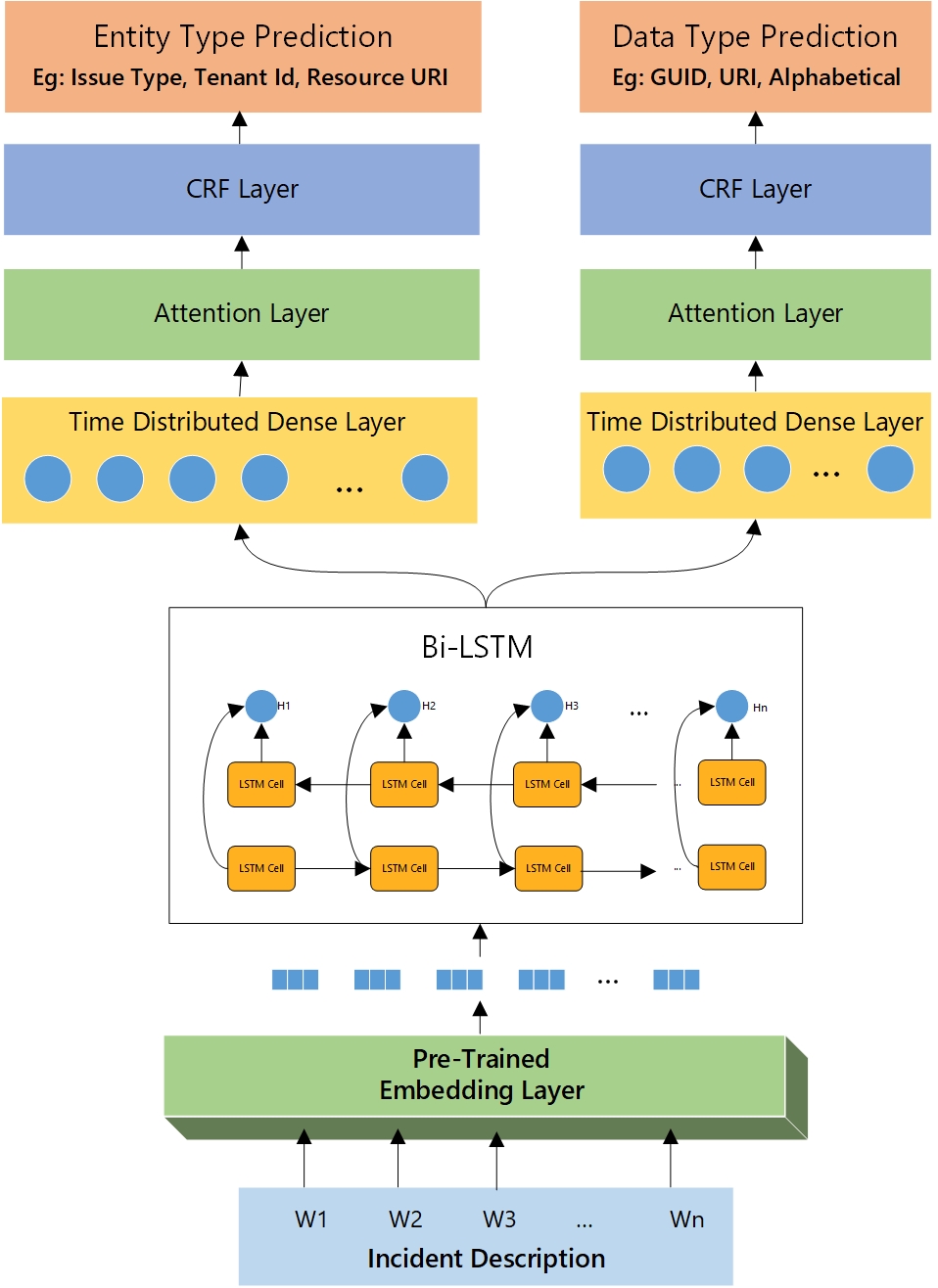}
\caption{\textbf{Multi-task model architecture}}
\label{fig:model-arch}
\end{figure}

\subsection{\textbf{Multi-Task Named-Entity Recognition Model}}
\label{subsubsec:multi-task-model}

The previous sections explain the phases of the \softner{} NER pipeline, as shown in Figure \ref{fig:pipeline}, that automate the significant task of creating labeled data. Here, we propose a novel Multi Task deep learning model that further generalizes entity extraction. The model solves two entity recognition tasks simultaneously - Entity Type recognition and Data Type recognition. The model uses an architecture, as described in Figure \ref{fig:model-arch}, that shares some common parameters and layers for both tasks, but also has task specific layers. Incident descriptions are converted to word level vectors using a pre-trained Glove Embedding layer. This sequence of vectors is interpreted by a Bi-directional LSTM layer. We then have distinct layers for the two tasks. The attention mechanism helps the model learn important sections of the sentences. Finally, the CRF layer produces a valid sequence of output labels. We perform back propagation using a combination of loss functions during training and evaluate tag level precision, recall, and F1 metrics. In the following sub sections we describe the important layers and approaches used in our model.

\smallskip
\subsubsection{Word Embeddings}
\label{subsubsec:glove-word-emb}

Language models in the semantic vector space, require real valued vectors as word representations \cite{collobert2011natural}. GloVe \cite{pennington2014glove} vectors, demonstrated on tasks such as word analogy and named entity recognition in \cite{pennington2014glove}, outperform various other word representations. Therefore, we use GloVe by creating an embedding layer with pre-trained GloVe weights loaded in our model as well as our baselines. 

\smallskip
\subsubsection{Bi-directional LSTM}
Recurrent Neural Networks (RNN) have been the basis for numerous language modelling tasks in the past \cite{mikolov2010recurrent}. But, RNNs tend to be biased towards more recent updates in long sequence scenarios. Long Short-term Memory (LSTM) networks \cite{hochreiter1997long} were designed to overcome the problems associated with vanilla RNNs. Their architecture allows them to capture long range dependencies using several gates. These gates control the portion of the input to give to the memory cell, and the portion from the previous hidden state to forget. Given a sentence as a sequence of real valued vectors $(x_1, x_2, .., x_n)$, the layer computes $\overrightarrow{h_{t}}$ which represents the leftward context of the word at the current time step $t$. A representation of a word receiving context from words occurring after it is achieved with a second LSTM that interprets the same sequence in reverse, returning $\overleftarrow{h_{t}}$ at each time step. This combination of forward and backward LSTM is referred to as Bi-Directional LSTM (BiLSTM) \cite{graves2005framewise}. The final representation of the word is produced by concatenating the left and right context, $h_{t}=[\overrightarrow{h_{t}};\overleftarrow{h_{t}}]$.

\smallskip
\subsubsection{Neural Attention Mechanism}
In recent years attention mechanism has become increasingly popular in various NLP applications like neural machine translation \cite{bahdanau2014neural}, sentiment classification \cite{chen2017recurrent} and parsing \cite{li2016discourse}. Novel architectures like transformers \cite{vaswani2017attention} and BERT \cite{devlin2018bert} have proven the effectiveness of such a mechanism for various downstream tasks. We implement attention at the word level as a neural layer, with a weight parameter $W_{a}$. It takes as input the the hidden states from the BiLSTM, transposed to output dimensions using a time distributed dense layer. Let $h = [h_1, h_2, .. h_{T}]$ be the input to the attention layer. The attention weights and final representation $h^*$ of the sentence is formed as follows:
\begin{equation}
    scores = W_a^Th
\end{equation}
\begin{equation}
    \alpha = softmax(scores)      
\end{equation}
\begin{equation}
    r = h\alpha^T    
\end{equation}
\begin{equation}
    h^* = \tanh(r)    
\end{equation}

\begin{figure}[h]
\includegraphics[width=1\linewidth]{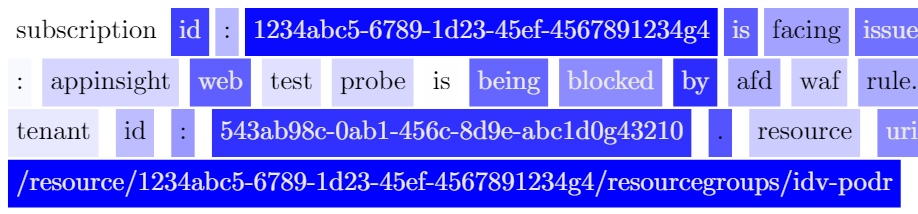}
\caption{\textbf{Attention visualization on a sample input}}
\label{fig:attention-viz}
\end{figure}

We visualize the attention vector $\alpha$ for a test sentence in Figure \ref{fig:attention-viz}, where we observe that it learns to give more emphasis to tokens that have a higher likelihood of being entities. In Figure \ref{fig:attention-viz}, the darkness in the shade of blue is proportional to the degree of attention. In case of long sequences, this weighted attention to certain sections of the sequence, that are more likely to contain entities, helps improve the model's recall/sensitivity.


\subsubsection{Conditional Random Fields}
Simply using hidden state representations ($h_t$) as word features to make independent tagging decisions at the word level leaves behind inherent dependencies across output labels in tasks like Named Entity Recognition. Our NER task also has this characteristic since the initial \softner{} heuristics enforce structural constraints, e.g. separators between key-value and html table tags. In learning these dependencies and generalizing them to sentences without these constraints, we model tagging decisions jointly using conditional random fields \cite{lafferty2001conditional}.

For an input sequence $\boldsymbol{X = (x_1, x_2, .., x_n)}$, let $\boldsymbol{y = (y_1, y_2, .., y_n)}$ a potential output sequence, where n is the no. of words in the sentence. Let $P$, the output of the BiLSTM network passed through the dense and attention layers, be the matrix of probability scores of shape $n \times k$, where k is the number of distinct tags. That is $P_{i,j}$ is a score that the $i^{th}$ word corresponds to the $j^{th}$ tag. We define CRF as a layer in the model, whose working is as follows.

\begin{equation}
    s(\boldsymbol{X,y}) = \sum_{i=0}^{n} A_{y_i,y_{i+1}} + \sum_{i=0}^{n} P_{i,y_{i}} 
\end{equation}
\begin{equation}
    p(\boldsymbol{y|X}) = 
    \dfrac{e^{s(\boldsymbol{X,y})}}{\sum_{y' \in Y} e^{s(\boldsymbol{X,y'})}  }
\end{equation}

Here A represents the matrix of transition scores where $A_{i,j}$ is the score for the transition from $tag_i$ to $tag_j$. The score $s(X,y)$ is converted to a probability for the sequence $y$ to be the right output using a softmax over $\boldsymbol{Y}$(all possible output sequences). The model learns by maximizing the log-probability of the correct $y$. While extracting the tags for the input, we predict the output sequence with the highest score - $y^* = argmax_{y' \in Y} p(y'|X)$.

\smallskip
\subsubsection{Multi-Task Learning}
Caruana et al. \cite{caruana1997multitask} defines Multi-Task Learning (MTL) as an approach to improve generalization in models by using underlying common information shared among related tasks. Some well known applications of MTL are multi-class and multi-label classification. In the context of classification or sequence labelling, MTL improves performance of individual tasks by learning them jointly.

In \softner{}, named-entity recognition is the primary task. In this task, models mainly learn from context words that support occurrences of entities. But we also observe that incorporating a complementary task of predicting the data-type of a token reinforces intuitive constraints indirectly on model training. For example in an input like ``\textit{The SourceIPAddress is 127.0.0.1}", the token \textit{127.0.0.1} is identified more accurately by our model, as the entity type \textit{Source Ip Address}, because it is also identified as the data-type \textit{Ip Address}, in parallel. This supplements the intuition that all \textit{Source Ip Addresses} are \textit{Ip Addresses}, thus, improving model performance.

As shown in Figure \ref{fig:model-arch}, we use a multi-head architecture, where the lower level features generated by the BiLSTM layers are shared, whereas the other layers are task specific. We define the entity type prediction as the main task and that of data type prediction as the auxiliary task. The losses are initially calculated individually for both tasks, $l_1$ and $l_2$, and then combined into $loss_{c} = \alpha \times l_1 + \beta \times l_2$. The parameter $loss\_weights = (\alpha,\beta)$ is used to control the importance between main and auxiliary tasks.
\section{Knowledge Graph Construction}
\label{sec:rel-extraction-knowledge-graphs}

Next, we describe our approach for mining entity relations and automatically constructing knowledge graphs.

\subsection{\textbf{Entity Relation Extraction}}
Once named-entities in incidents are tagged by \softner{}'s trained NER model, we recognize pairs of related entities, that is, binary relations. Here, instead of directly classifying all possible relation instances ($n$-ary), we first identify whether a given pair of entities is related or not. With the aim of building an unsupervised framework, similar to prior work \cite{mcdonald2005simple}, \cite{zelenko2003kernel}, we use co-occurrence of entity pairs in a sentence to extract binary relations. By extracting all possible entity pairs that follow this assumption, we get a noisy candidate set of binary entity relations based solely on co-occurrence.

Consequently, we then score each candidate tuple using a co-occurrence based measure and filter noisy candidates. In information theory, \textit{mutual information} (MI) \cite{fano1961transmission} of 2 random variables is a measure of the ``amount of information" obtained about one variable through observing the other. While \textit{mutual information} averages the measure over all possible outcomes, \textit{pointwise mutual information} (PMI) \cite{church-hanks-1990-word} is defined for a single event (i.e. pair of outcomes). Mathematically, PMI is defined as described in equation \ref{eq:pmi}.

\begin{equation}
\label{eq:pmi}
    \operatorname {pmi} (x;y)\equiv \log {\frac {p(x,y)}{p(x)p(y)}}=\log {\frac {p(x|y)}{p(x)}}=\log {\frac {p(y|x)}{p(y)}}
\end{equation}
\begin{equation}
\label{eq:symmetric}
    \operatorname {pmi} (x;y)=\operatorname {pmi} (y;x)
\end{equation}

PMI has been applied for finding collocations and associations between tokens \cite{church-hanks-1990-word}, \cite{thanopoulos2002comparative} by leveraging frequency of occurrences to approximate probabilities. We observe that our aim to score co-occurring entity pairs on their relatedness is analogous to these applications. Thus, we use a variant of PMI - \textbf{\textit{normalized pointwise mutual information} (NPMI)} \cite{bouma2009normalized}, to score each entity pair - $(e_1,e_2)$, as described in equations \ref{eq:npmi-start}-\ref{eq:npmi-end}.

\begin{equation}
\label{eq:npmi-start}
    \operatorname {npmi} (e_1;e_2)={\frac {\operatorname {pmi} (e_1;e_2)}{-\log {p(e_1,e_2)}}}
\end{equation}

\begin{equation}
    \operatorname {pmi} (e_1;e_2)=\log {\frac {p(e_1,e_2)}{p(e_1)p(e_2)}}
\end{equation}

\begin{equation}
\label{eq:npmi-end}
    p(e_1) = \frac{f_1}{f_{total}}; \hspace{0.5em} p(e_2) = \frac{f_2}{f_{total}};  \hspace{0.5em} p(e_1,e_2) = \frac{f_{joint}}{f_{total}};
\end{equation}

In the above equations and in Table \ref{tab:binary-relation-examples}, $f_1$ \& $f_2$ are frequencies of the entities $e_1$ \& $e_2$ respectively, $f_{joint}$ is the frequency of co-occurrence of the entity pair, and $f_{total}$ is the total frequency of all entities. Note that, from equation \ref{eq:symmetric}, PMI, and consequently NPMI is symmetric. Therefore, all binary entity relations extracted are undirected in nature. NPMI scores $\in [-1,1]$, resulting in -1 for never occurring together, 0 for independence, and +1 for complete co-occurrence. Using this, we eliminate all entity pairs with NPMI $<$ 0 as noise, leaving us with a final set of binary entity relations. Table \ref{tab:binary-relation-examples} shows some examples of entity pairs and their NPMI scores.

\begin{table}[h]
\centering
\caption{\textbf{Examples of extracted Binary Entity Relations}}
\label{tab:binary-relation-examples}
\setlength{\tabcolsep}{2pt}
\small
\begin{tabular}[t]{l c r} 
\hline
\textbf{Entity Pair} & $\boldsymbol{(f_1,f_2,f_{joint})}$ & \textbf{NPMI}\\
\hline
(\textit{Remote Port Range, Remote Address}) & (564, 558, 557) & 0.99 \\
(\textit{Tunnel Name, Encap Type}) & (761, 654, 486) & 0.88\\
(\textit{VNet Name, VNet Id}) &  (860, 985, 432) & 0.78 \\
(\textit{Gateway Id, VNet Id}) & (1071, 985, 124) & 0.47 \\
(\textit{Tunnel Name, Destination IP}) & (761, 72908, 1) & -0.38 \\
(\textit{Destination IP, Subscription Name}) & (72908, 19125, 11) & -0.55 \\
\hline
\end{tabular}
\end{table}

\begin{figure}[h]
\centering

\begin{tikzpicture}[
roundnode/.style={circle, draw=black!80, fill=green!20, very thick, minimum size=0.5mm}
]
	\begin{pgfonlayer}{nodelayer}
		\node [style=roundnode, label={left:\textit{VNet Id}}] (7) at (-8.5, 1) {};
		\node [style=roundnode, label={left:\textit{VNet Name}}] (8) at (-8.5, -1) {};
		\node [style=roundnode, label={below:\textit{VNet Region}}] (9) at (-5, -2) {};
		\node [style=roundnode, label={below:\textit{Tunnel Name}}] (11) at (0.5, -1.5) {};
		\node [style=roundnode, label={above:\textit{Encap type}}] (12) at (3.25, 0) {};
		\node [style=roundnode, label={above:\textit{Subscription Id}}] (14) at (-5, 2) {};
		\node [style=roundnode, label={above:\textit{Gateway Id}}] (15) at (-2, 0) {};
	\end{pgfonlayer}
	\begin{pgfonlayer}{edgelayer}
		\draw (7.center) to (15.center);
		\draw (7.center) to (9.center);
		\draw (7.center) to (8.center);
		\draw (14.center) to (7.center);
		\draw (8.center) to (9.center);
		\draw (15.center) to (11.center);
		\draw (11.center) to (12.center);
		\draw (8.center) to (15.center);
		\draw (14.center) to (8.center);
	\end{pgfonlayer}
\end{tikzpicture}
\caption{\textbf{Sub-graph of Related Entities}}
\label{fig:sub-graph}
\end{figure}
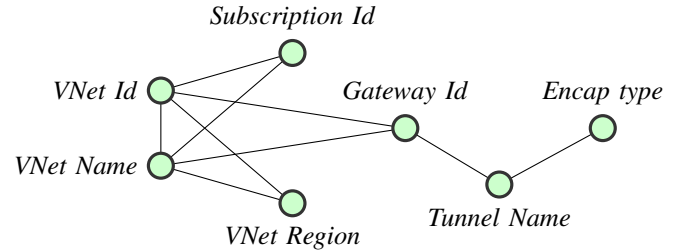

\subsection{\textbf{Entity Knowledge Graph}}

A knowledge graph formally represents semantics by describing entities and relationships. It captures information that can be queried like traditional databases, analyzed like graph data structures, and allows inference of new knowledge. In the incident management space, this unlocks various applications, such as mining complex interactions of cloud resources (Table \ref{tab:complex-relation-examples}), or inferring the relevance of an entity to the issue described in an incident (Section \ref{subsec:entity-incident-relevance}).

Having identified binary relations, we construct an undirected knowledge graph $\boldsymbol{G=(V,E)}$, where nodes $\boldsymbol{V}$ are entities and edges $\boldsymbol{E}$ are binary relations between pairs of entities. We also assign weights $\boldsymbol{W_{e_i,e_j} = \operatorname {npmi} (e_i;e_j)}$, to the edges between all entity pairs ($\boldsymbol{e_i$,$e_j}$). Figure \ref{fig:sub-graph} shows a sub-graph of related entity types in the entity knowledge graph constructed from our incident data set. These relations are utilized to construct the complete knowledge graph, as shown in Figure \ref{fig:kgraph-overview}.

\begin{table*}
\centering
\caption{\textbf{Examples of extracted Complex ($n$-ary) Entity Relations}}
\label{tab:complex-relation-examples}
\begin{tabular}{ll}
\hline
\textbf{Related Entities} & \textbf{Relation Description}\\
\hline
(\textit{Destination IP, AS path, Output packets}) & Describing a BGP routing incident causing connection issues. \\[-1pt]

(\textit{Correlation Id, Allocation Id, VNet Id, MAC Address, Container Id}) & Describing various tasks in network manager setup for a VM. \\[-1pt]

(\textit{VNet Name, VNet Id, Gateway Id, Tunnel Name, Encap Type}) & Describing a VNet gateway instance that is down. \\[-1pt]

(\textit{Error Code, Error message, is retriable exception, is user error}) & Entities that describe errors and exceptions. \\[-1pt]

(\textit{VNet Name, VNet id, VNet region, v net name, v net id, Resource URI}) & Entities co-referring a single resource related to the incident. \\[-1pt]

\hline
\end{tabular}
\end{table*}

We briefly explore mining more complex $n$-ary relations from our knowledge graph. One simple approach is to view $n$-ary relations as graph cliques - a subset of vertices of an undirected graph such that every pair of vertices are adjacent (pairwise related entities) \cite{mcdonald2005simple}. To overcome overlaps in cliques, we query maximal-cliques, that is, those cliques that are not subsets of other cliques. Table \ref{tab:complex-relation-examples} shows some examples of complex relations extracted with this method and what they describe. While we do not perform statistical analysis, examples from Table \ref{tab:complex-relation-examples} suggest that the constructed knowledge graph is able to fairly represent complex relationships by factoring them as binary relations.
\section{Implementation}
\label{sec:implementation}
The \softner{} implementation and deployment comprises of various modules. First, we train the ML models using historical incident data from Microsoft's services. Next, we deploy the models to a scalable REST API using the Flask web framework. Lastly, we integrate the \softner{} API into the incident management system at Microsoft. Here, we describe each of these in detail:

\subsection{\textbf{Model training}}
We have implemented \softner{} and all the machine learning models using Python 3.7.5, with Keras-2.2.4 and the tensorflow-1.15.0 backend. The hyper-parameters for the deep learning models are set as follows: word embedding dimension used is 100, hidden LSTM layer size is set to 200 cells, and, maximum length of a sequence is limited to 300. These optimal  hyper-parameters were chosen to train a robust yet light weight model and were re-used among all models. The embedding layer uses pre-trained weights from  stanford-nlp's glove.6B.100d. Our models are trained on an Ubuntu 16.04 LTS machine, with 24-core Intel Xeon E5-2690 v3 CPU (2.60GHz), 112 GB memory and 64-bit operating system. The machine also has a Nvidia Tesla P100 GPU with 16 GB RAM.

\subsection{\textbf{Model deployment}}
We have also deployed \softner{} as a REST API developed using the Python Flask web app framework. The REST API offers a POST endpoint which takes the incident description as input and returns the extracted entities in JSON format. We have deployed it on Microsoft's Azure cloud platform which allows us to automatically scale the service based on the variation in request volume. This enables the service to be cost efficient since majority of the incidents are created during the day. We have also enabled application monitoring which alerts us in case the availability or the latency regresses.

\subsection{\textbf{Integration}}
At Microsoft, we have thousands of production services built and operated by tens of thousands of developers. In order to effectively and efficiently handle the issues and regressions in these services, we have a dedicated incident management platform called IcM. This platform allows internal and external partners to create an incident against various services and teams. The IcM platform provides an extensibility mechanism using which custom modules can be enabled which can subscribe to various events such as incident creation, update, mitigate and resolution. We have integrated \softner{} API with the IcM platform to surface the insights directly into the IcM portal. We have enabled multiple integration points based on the user scenario:
\begin{itemize}
    \item \textit{Manual Trigger} - Any developer can enter an incident Id and see the \softner{} results in real time. This is useful when a team or a developer wants to try out \softner{}.
    \item \textit{Auto Trigger} - Service owners can also enable \softner{} for the incidents belonging to their respective teams. This way, the insights are added to the incident automatically even before an on-call engineer is engaged.
    \item \textit{Integration with other modules} - The entities extracted using \softner{} can be used to trigger other diagnostic modules. For example, a network diagnostic module may require the VNet Id and the Source IP Address to localize a given incident. \softner{} is able to extract these entities automatically and can feed them to such diagnostic modules.
\end{itemize}

\section{Evaluation}
\label{sec:evaluation}

\softner{} solves the problem of knowledge extraction from unstructured text descriptions of incidents. To evaluate the \softner{} framework in its entirety, we propose a three phase evaluation:
\begin{itemize}
    \item \textbf{Entity Types}: How does \softner{}'s unsupervised approach perform in recognizing distinct entity types?
    \item \textbf{Named-Entity Recognition}: How does \softner{}'s Multi-Task model compare to state-of-the-art deep learning approaches for the NER task?
    \item \textbf{Entity Relations}: How does \softner{}'s unsupervised approach perform in extracting and scoring entity relations using NPMI?
\end{itemize}

\subsection{\textbf{Study Data}}
\label{subsec:study-data}

In the following evaluation experiments, we apply \softner{} to service incidents at \CompanyX{}, a major cloud service provider. These are incidents retrieved from large scale online service systems, which have been used by a wide distribution of users. In particular, we collected incidents spanning over a time period of 2 months. Each incident is described by its unique Id, title, description, last-modified date, owning team name and, also, whether the incident was resolved or not. Incident description is the unstructured text with an average of 472 words, showing us how verbose the incident descriptions are. Owning Team Name here, refers to the team to which the incident has been assigned.

\subsection{\textbf{Entity Type Evaluation}}
\label{subsec:entity_type_evaluation}

Here, we evaluate the effectiveness of \softner{}'s unsupervised approach for named-entity extraction. Specifically, we evaluate the correctness of the entity types extracted by \softner{} on the entire study data. As the component performs \textbf{unsupervised} entity extraction, we manually evaluate the precision of extraction. Top 100 (limited to 100 since evaluation is manual) most frequent distinct entities are extracted by the component. We then, manually validate each potential entity and analyse the precision that is the fraction of extracted entities that are actually software entities.

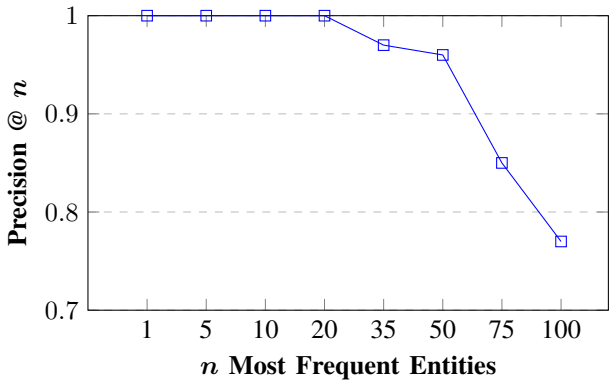
\begin{figure}[h]
\centering
\begin{tikzpicture}
\begin{axis}[
    height=5.5cm,
    width=8.5cm,
    xlabel={\textbf{$\boldsymbol{n}$ Most Frequent Entities}},
    ylabel={\textbf{Precision @ $\boldsymbol{n}$}},
    xmin=0,
    ymin=0.7, ymax=1.0,
    xtick = {1,2,3,4,5,6,7,8},
    xticklabels={1, 5, 10, 20, 35, 50, 75, 100},
    ytick={0.7, 0.8, 0.9, 1.0},
    legend pos=south west,
    ymajorgrids=true,
    grid style=dashed
]

\addplot[color=blue,mark=square]
    coordinates {
    (1,1.0) (2,1.0) (3,1.0) (4,1.0) (5,0.97)
    (6,0.96) (7,0.85) (8,0.77)
    };
    
\end{axis}
\end{tikzpicture}
\caption{\textbf{Precision vs Rank curve for Entity Types}}
\label{pattern-extraction}
\end{figure}

Since the precision of \softner{}'s entity type extraction depends on the frequency of occurrence of entities, we further plot precision against a cut off rank $n$. Figure \ref{pattern-extraction} summarizes the precision of \softner{}'s entity type extraction against the top $n$ entities extracted, where $n \in [1,100]$. From this analysis, we see that \softner{} is able to extract 77 valid entities per 100 entities. In this experiment, $n$ corresponds to the rank of the entity extracted with respect to frequency of occurrence. That is, a higher $n$ refers to an entity with low frequency of occurrence, which in turn can be extrapolated as an entity that is less important. We thus see an expected decrease in precision, as $n$ increases, due to noisy tokens (false positives) like \textit{"to troubleshoot issue"} and \textit{"for cleanup delay"}. \softner{}'s unsupervised entity type extraction has a minimal precision variation, also known as fall out rate, of 0.23 for an $n$ value as high as 100. This strengthens the hypothesis that \softner{}'s pattern extractors can pick up entities from unstructured text effectively, in a completely unsupervised manner.

\begin{table}[h]
\small
\centering
\caption{\textbf{NER Model Evaluation}}
\vspace{-1pt}
\label{tab:model-evaluation}
 \begin{tabular}[t]{lccc} 
\hline
 \textbf{Metric} & \textbf{BiLSTM-CRF} & \textbf{BiLSTM-CRF} & \textbf{SoftNER} \\
  &  & \textbf{ Attention} & \textbf{Model} \\
 \hline
 Avg F1 & 0.8803 & 0.8822 & \textbf{0.9572} \\
 Weighted Avg F1 &  0.9401 & 0.9440 & \textbf{0.9682} \\
 Avg Precision & 0.9160 & 0.9088 & \textbf{0.9693} \\
 Avg Recall &  0.8669 & 0.8764 & \textbf{0.9525} \\
 \hline
\end{tabular}
\end{table}

\subsection{\textbf{Named-Entity Recognition Evaluation}}
Here, we evaluate the SoftNER deep learning model on the Named-Entity Recognition Task. We compare the multi task model, described in section \ref{subsubsec:multi-task-model} and Figure \ref{fig:model-arch}, against two baseline models, BiLSTM-CRF and BiLSTM-CRF with attention mechanism. These baseline models are state-of-the-art for NER \cite{huang2015bidirectional}, \cite{lample2016neural}, \cite{chiu2016named} and other NLP tasks as well. The models are compared on a fixed test set that accounts for $20\%$ of the ground-truth data set labeled using the unsupervised approach as described in \ref{subsec:unsupervised_data_labelling}. Note that we also ensure the incidents in the test set occur after those in the training set temporally. We use average precision, recall and F1 metrics to evaluate and compare the models on the NER task. The metrics are averaged over the distinct entities types tagged by the model.

As shown in Table \ref{tab:model-evaluation}, we observe that the baseline BiLSTM-CRF, with and without attention mechanism, achieves an average F1 score of around 0.88. Whereas, \softner{}'s Multi Task Model, as described in Section \ref{subsubsec:multi-task-model}, achieves a higher average F1 score of around 0.96, i.e., a $\Delta F1\%$ of 8.7\%. We also observe a high average recall of 0.95, reflective of a robust ability to extract a lot of relevant information from descriptions which directly correlates with the ease of understanding the problem and identifying resources affected by the incident.

We further analyze the generalization of the model by analyzing test examples that were falsely labeled. Table \ref{tab:false-examples} shows a few examples of sentences and the entities extracted from them. Note that we refer to false positives as FP, and, false negatives as FN in the table. We observe that some of the FPs are actually correct and were mislabeled in the test set because of the limitations of the pattern extractors. Let's take Example 1 from Table \ref{tab:false-examples} for instance. Here, the unsupervised labelling component was only able to label \textbf{\textit{``2aa3abc0-7986-1abc-a98b-443fd7245e6"}} as \textbf{Subscription Id}, but not \textbf{\textit{``vaopn-uk-vnet-sc"}} as \textbf{Vnet Name}, due to restrictions with pattern extractors and label propagation. But the \softner{} model was able to extract both the entities from the sentence, proving it's ability to generalize beyond obvious structural pattern rules visible in the training data. Row 2 shows a similar false positive example with the extraction of \textbf{\textit{192.168.0.5}} as \textbf{IP Address}. We also show a few contrasting false negatives, in rows 4 and 5, where the model was unable to extract entities \textbf{Ask} and \textbf{Ip Address} respectively. 

\begin{table}
\centering
\caption{\textbf{FP and FN examples}}
\label{tab:false-examples}
\begin{tabular}[t]{p{4cm} p{1cm} p{2.5cm}} 
\hline
\textbf{Sentence} & \textbf{Result} & \textbf{Entities Tagged}\\
\hline
\textit{SubscriptionId:2aa3abc0-7986-1abc-a98b-443fd7245e6 unable to delete vnet name vaopn-uk-vnet-sc} & FP & {2aa3abc0-7986-1abc-a98b-443fd7245e6, vaopn-uk-vnet-sc }\\

\textit{Device Name: njb02-23gmk, pa: 192.168.0.5 could not be configured!} & FP & {njb02-23gmk, 192.168.0.5} \\

\textit{The customer's  main ask: Need help to access cloud storage}  & FN & - \\

\textit{The loopback (ipv4) address (primary) is 192.131.75.235} & FN & {ipv4} \\
\hline
\end{tabular}
\end{table}

\subsection{\textbf{Entity Relation Evaluation}}
\begin{figure}[h]
\centering
\begin{tikzpicture}
\begin{axis}[
    height=5.5cm,
    width=8.5cm,
    xlabel={\textbf{$\boldsymbol{n}$ Most Frequent Entity Pairs}},
    ylabel={\textbf{Precision @ $\boldsymbol{n}$}},
    xmin=0, xmax=150,
    ymin=0.7, ymax=1.0,
    xtick={1, 17,  33,  49,  65,  81,  97,  113, 129, 145},
    ytick={0.6, 0.7, 0.8, 0.9, 1.0},
    legend pos=south east,
    ymajorgrids=true,
    grid style=dashed,
]

\addplot[color=blue,mark=square]
    coordinates {
    (1,1.0) (9,1.0) (17,1.0) (25,1.0) (33,1.0) (41,1.0) (49,1.0) (57,0.9649) (65,0.9692) (73,0.9589) (81,0.9506) (89,0.9550) (97,0.9072) (105,0.8857) (113,0.8761) (121,0.8842) (129,0.8604) 
    (137,0.8394) (145,0.8344) (150,0.8333)
    };
    
    
\legend{Precision}
\end{axis}
\end{tikzpicture}
\caption{\textbf{Precision vs Rank for Entity Relations}}
\label{fig:npmi-eval}
\end{figure}
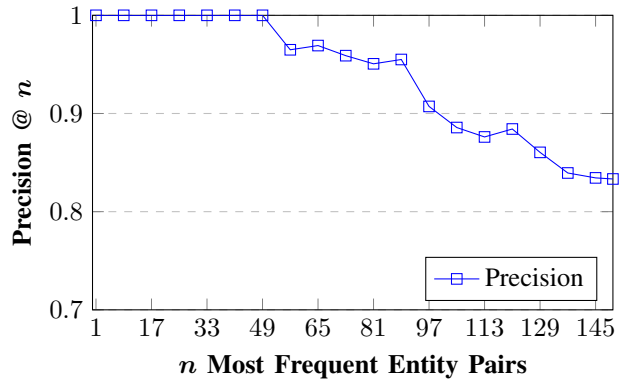

Next, we evaluate our unsupervised approach to extract binary entity relations and score them using NPMI. These binary entity relations and their scores represent the crux of the knowledge graph constructed in Section \ref{sec:rel-extraction-knowledge-graphs}. In turn, any inference performed on the graph, like entity recommendation (Section \ref{subsec:entity-incident-relevance}), depends on the quality of these relations. Consequently, we manually evaluate the precision of NPMI scores given to each entity pair. First, the top 150 (limited to 150 since evaluation is manual) most frequent entity pairs are sampled. We then manually validate the correctness of NPMI computed for each entity pair and analyze its precision, i.e. the fraction of scores that are valid. We further plot precision against rank $n \in [0,150]$ in Figure \ref{fig:npmi-eval}. Here, $n$ corresponds to the rank of the entity pair extracted with respect to frequency of co-occurrence ($f_{joint}$). From this analysis, we see that NPMI scores are valid for 90 per 100 entity pairs. A precision of 0.9 for a $n$ value as high as 100, strengthens the hypothesis that binary entity relations can be effectively extracted and scored with our approach.

\section{Applications}
\label{sec:applications}

Automated knowledge extraction from service incidents can unlock several applications and scenarios. Here, we explore and evaluate the value of extracted knowledge for two applications. First, we show that entities extracted by \softner{} can be utilized to improve simple machine learning models for incident triaging. Next, we show that the knowledge graph can be used to build entity recommenders that can improve tooling in incident management platforms and in turn reduce customer impact.

\subsection{\textbf{Auto-Triaging of Incidents}}
Incident triaging is the process of assigning a new incident to the responsible team. This is currently manually performed by on-call engineers. It is not uncommon for an incident to be rerouted to different teams until the appropriate team is engaged, thereby reducing the accuracy and efficiency of incident management. Based on an empirical study, Chen et al. \cite{EmpiricalIcMICSE2019} showed that the reassignment rate for incidents can be as high as 91.58\% for online services at Microsoft. Several efforts \cite{EmpiricalIcMICSE2019}, \cite{ContinuousTriageASE2019} have been made to automate the triaging process by leveraging the title, description and other meta-data of the incidents. Here, we evaluate the effectiveness of the knowledge extracted by \softner{} for the downstream task of automated triaging of incidents. Incident triaging is essentially a multi-class classification problem since the incident could be assigned to one of many teams.

\begin{table}[h]
\small
    \begin{center}
    \caption{\textbf{Comparison of Accuracy for Auto-Triaging}}
    \label{tab:auto-triaging-models}
    \setlength\tabcolsep{2 pt}
    \begin{tabular}{@{}p{2.66cm}ccccc@{}}
        \hline
        \textbf{Feature Set}  & \textbf{Random} & \textbf{Linear} & \textbf{Gaussian} &  \textbf{K-Nearest} & \textbf{Naive} \\[-2pt]
         & \textbf{Forest} & \textbf{SVM} & \textbf{SVM} & \textbf{Neighbors} & \textbf{Bayes}\\
        \hline
        Title + Description & 74.64 & 85.93 & 87.06 & 81.32 & 69.69 \\
        \hline
        \softner{} Entities & 93.38 & 93.34 & 93.39 & 92.40 & 87.67\\
        $\Delta$ \% & 22.31 & 8.26 & 7.02 & 12.76 & 22.85 \\
        \hline
        \softner{} Entities + Title & \textbf{98.60} & \textbf{99.20} & \textbf{98.95} & \textbf{99.14} & \textbf{88.07} \\
        $\Delta$ \% & 27.66 & 14.34 & 12.78 & 19.75 & 23.30 \\
        \hline
    \end{tabular}
    \end{center}
\end{table}

We sample 20\% of resolved incidents for the $10$ most common teams from the initial incident set (refer Section \ref{subsec:study-data}) and run the \softner{} model on the description to extract entities. These extracted entity values can now act as additional features to triaging models. The \softner{} entities can be broadly classified as either categorical or descriptive. While the descriptive entities are transformed to word embeddings using the same process described in Section \ref{subsubsec:glove-word-emb}, the categorical entities are encoded into one-hot vectors. We then look at different combinations of features and compare the 5-fold cross-validation accuracy on various classification models. It is evident from Table \ref{tab:auto-triaging-models} that the models using the \softner{} entities as features, either on their own or along with the title, outperform the baselines that use only raw title and description information. We observe significant margins, with up to $7$\% - $27$\% increase in the cross-validation accuracies. These results reinforce that the entities extracted by \softner{} are indeed useful and can significantly help in downstream tasks. Using the entities extracted by \softner{} also reduces the input feature space since we no longer have to use the whole incident description. We also achieve high performance using simple machine learning models thereby eliminating the need for complex deep learning models which have proven to be superior in past studies \cite{EmpiricalIcMICSE2019}.

\begin{table}[h]
\small\centering
    \caption{\textbf{Importance scores for top features}}
    \label{tab:feature_significance}
    \setlength\tabcolsep{2 pt}
    \begin{tabular}{lccccc}
        \hline
        \textbf{Feature}  & Exception & Problem & Ask &  Issue & Title \\[-3pt]
        & Message & Type & & \\
        \hline
        \textbf{Importance} & 0.0133 & 0.0111 & 0.0097 & 0.0051 & 0.0009\\
        \hline
    \end{tabular}
\end{table}

In addition to comparing accuracy, we analysed feature significance by using the feature\_importances\_ attribute of a random forest model trained on the various input features. As shown in Table \ref{tab:feature_significance}, we observe that the entities extracted by \softner{} were given more importance compared to the `Title', with top features being - `exception message', `problem type', `ask' and `issue'. This re-emphasises that the entities extracted from \softner{} boost the performance of classification models for the downstream task of automatic triaging of incidents.

\subsection{\textbf{Entity-Incident Relevance and Recommendation}}
\label{subsec:entity-incident-relevance}

Although extracting all entities from an incident is useful, certain entities are more important for the investigation and mitigation of an incident. Let’s consider a real incident reported by a customer
of the Cloud Networking service operated by \CompanyX{}. The incident was caused due to an error in deleting a Virtual Network resource. The key information required to mitigate this incident is actually the \textbf{\textit{Virtual Network (VNet) Id}}. But the customer had not mentioned the \textit{VNet Id} in their description. In this case, it would be useful to either intelligently recommend or incorporate such entities as mandatory in incident reporting forms to alleviate downstream mitigation tasks. Thus, here we evaluate the effectiveness of the knowledge extracted by SoftNER to infer the relevance of entities to the issue described in the incident for recommendation.

Previously, statistical entity-topic models \cite{newman2006statistical}, \cite{kim2012etm} have been studied to map named-entities to topics for document topic analysis. Bhargava et al. \cite{bhargava2019learning} proposed multiple methods to learn to map wikidata entities to pre-defined topics. The challenge, in our case, was the lack of a labeled set of incidents based on pre-defined topics. Also, manually identifying topics apriori would be difficult to scale to a multitude of services. As a result, in our approach, we use clustered incident titles as representatives of topics and map a subset of ranked related entities leveraging the previously constructed knowledge-graph.

\smallskip
\subsubsection{Clustering Incident Titles}
Inferring a subset of entities for each incident individually poses challenges. It is difficult to apply rule-based approaches since they are computationally expensive and also cannot be generalized to unseen incident examples. Hence, we first aim to group incidents by clustering their titles, which are generally a representative summary of the incident. Here, we convert incident titles to 100 dimension embeddings by averaging GloVe \cite{pennington2014glove} vectors for all tokens in the title. For clustering, we make use of DBSCAN - Density-Based Spatial Clustering of Applications with Noise \cite{ester1996density}, a clustering algorithm that does not require us to define the number of clusters apriori. The DBSCAN hyper-parameter $\boldsymbol\epsilon$, i.e. maximum distance between two samples for them to be in the same neighborhood, was tuned using the elbow method and plotting k-distance graphs, as suggested in the original paper \cite{ester1996density}. This provided us with over $50$ clusters on our incident data set. From Table \ref{tab:entity-titles-mapping}, we observe that clustering reduces complexity and also uncovers underlying topics in incidents.

\smallskip
\subsubsection{Inferring Related Entities}
Having clustered incident titles, and as a result the corresponding incidents, we now infer a related entity set for each cluster. First, for each cluster, we extract the top-5 most frequently occurring entities in the cluster's incidents (ranked by frequency). Next, we loop over this top-5 list and search for the first entity that exists in the knowledge graph $G=(V,E)$ constructed previously. Let this entity be called the ``\textbf{\textit{primary entity}}" ($\boldsymbol{e_p}$) of a cluster $C$. We then use the \textit{primary entity} as the source node to find the shortest paths to every other reachable entity. 
We hypothesize that every entity ($\boldsymbol{e_x}$) reachable from the primary entity ($\boldsymbol{e_p}$) of a cluster is related to the incidents of that cluster. Then, the relatedness between any reachable entity ($\boldsymbol{e_x}$) and a cluster ($C$) is scored as the average of the edge weights in the path taken from the \textit{primary entity} to the entity of interest ($\boldsymbol{e_x}$). Mathematically, it is defined as stated in equation \ref{eq:relatedness-score}. We then rank these entities based on the score computed and choose the top-K entities as entities related to the cluster. Note, that the primary entity itself is given a score of $1.0$. Also, we use averaging instead of product here, since the edge weights $\boldsymbol{W_{e_i,e_j} = \operatorname {npmi} (e_i;e_j)}$, and NPMI is a function of log probabilities (Section \ref{sec:rel-extraction-knowledge-graphs}).

\begin{equation}
\label{eq:relatedness-score}
\begin{split}
    \operatorname {relation-score} (e_x;C)=
    {
    \boldsymbol{
        \frac{
            \sum_{\forall e_i \in path(e_p, e_x)} W_{e_i,e_{i+1}}}
            {\sum_{\forall e' \in path(e_p, e_x)} 1}
        }
    } \\
    \\
    \text{where } path(\boldsymbol{e_p}, \boldsymbol{e_x}) = \text{shortest path from } \boldsymbol{e_p} \text{ to } \boldsymbol{e_x} 
\end{split}
\end{equation}

With the above approach, top-$K$ related entities for each cluster are pre-determined and stored. For a newly created incident with a complete incident title, first, the nearest cluster for that title is found. Then, the pre-determined top-$K$ entities for that cluster are recommended for that incident.

Table \ref{tab:entity-titles-mapping} shows a few examples of clustered titles and top-5 related entities extracted using the approach described above. While we leave more formal statistical evaluation of clustering for future work, we make some informal observations here, inferring from examples in Table \ref{tab:entity-titles-mapping}. Let's take example 1 for instance. Here, the clustered titles suggest incidents where the customer is \textbf{\textit{unable to deleted a virtual network}}. From the top-5 related entities, we see that our approach correctly identifies key entities, such as \textbf{\textit{Vnet Id}} and \textbf{\textit{Vnet Name}}, that point to the resource of interest, i.e. the virtual network. We also observe that it identifies the \textbf{\textit{Allocator Service}} responsible for allocation/de-allocation of the virtual network. Example 2 shows a cluster of incidents that describe \textbf{\textit{aggregated issues in the control path}} of a virtual resource. In this case, an entire hierarchy of entities, from \textbf{\textit{Subscription}} to \textbf{\textit{Deployment}} to an individual \textbf{\textit{Resource}}, is identified. These recommendations are effective to quickly identify resources affected, layers of dependencies, and mitigation steps to reduce customer impact.

\begin{table}
\centering
\caption{\textbf{Titles and Top-5 Related Entities}}
\label{tab:entity-titles-mapping}
\begin{tabular}[H]{p{4.5cm} p{3.5cm}} 
\hline
\thead[cl]{\textbf{Clustered Titles}} & \thead[cl]{\textbf{Top-5 Related Entities}}\\
\hline
\makecell[cl]{\textit{Unable to delete Vnet }
\\ \textit{Vnet stuck in updating state} 
\\ \textit{Vnet: Unable to delete Public IP}} & 
\makecell[cl]{(VNet Id, $1.0$) 
\\ (Allocator Service URI, $0.70$)
\\ (Availability Zone, $0.68$)
\\ (VNet Name, $0.66$)
\\ (Tunnel Name, $0.63$)}\\[2.5em]

\makecell[cl]{\textit{VM Network Profile fails to load} 
\\ \textit{High IO latency networking suspicious} 
\\ \textit{Ping Mesh Drops networking suspicious}} & 
\makecell[cl]{(Subscription Id, $1.0$) 
\\ (VNet Name, $0.41$) 
\\ (Deployment Id, $0.39$) 
\\ (Resource Id, $0.35$) 
\\ (VNet Id, $0.33$)}\\[2.5em]

\makecell[cl]{\textit{MAC Request - Port Provisioning }
\\ \textit{Port provisioning 1 device}
\\ \textit{Ports mop sent for review}} & 
\makecell[cl]{(Closed in SLA, $1.0$)
\\ (SLA w/o Dependency, $0.44$) 
\\ (Device Count, $0.40$)
\\ (Dependency Time, $0.37$) 
\\ (Allocator Service URI, $0.31$)}\\[2.5em]

\makecell[cl]{\textit{Cluster unhealthy - Interface Down} 
\\ \textit{Unhealthy cluster - BGP Session Down} 
\\ \textit{Cluster is unhealthy : Device Reloaded}} & 
\makecell[cl]{(SNMP Interface Index, $1.0$)
\\ (Allocator Service URI, $0.32$) 
\\ (Availability Zone , $0.31$)
\\ (Resource Id, $0.28$)
\\ (Interface Address, $0.27$)}\\
\hline
\end{tabular}
\end{table}

\section{Discussion}
\label{sec:discussion}
In this section, we first discuss the generalizability of
SoftNER and then discuss some ideas for future work.

\subsection{\textbf{Generalizing \softner{}}}
Today, incident management in large-scale cloud services is largely a manual process. On-call engineers have to read through the incident reports, extract the relevant information and then do root cause analysis and mitigation. This manual intervention also causes downstream impact on service reliability and customer satisfaction. Manual understanding and parsing of incident reports is a key bottleneck for incident automation. In this work, we have designed and deployed \softner{} at Microsoft for knowledge mining from incident reports. However, the problem and applications are generic and applicable to any cloud service provider. Further, since we have built \softner{} in an extensible manner using unsupervised techniques, it can be trained and applied to any service. We bootstrap the knowledge graph using syntactic patterns such as key-value pairs and tables which are generic and agnostic to the service. New patterns can also be integrated very easily into the machine learning pipeline, if required.

\subsection{\textbf{Future Wor}k}
\softner{} currently mines knowledge graphs from incidents by extracting named-entities and inferring relations between the entities. As next steps, we plan to expand the knowledge mining and leverage it for more scenarios:

\begin{enumerate}
    \item \textbf{Incident summarization} - Lot of critical information about the incidents is embedded in natural language. For instance, incident details like symptoms, reproduction and mitigation steps is described using natural language. So, we will extend the knowledge mining beyond named-entities to even parsing and understanding natural language. The extracted information can then be used for several applications such as incident summarization.
    
    \item \textbf{Automated health-checks} - Often times, on-call engineers have to pull up telemetry and logs for the resources affected by an incident. Or, they might look up the resource allocation for a given subscription. We will integrate \softner{} with the log mining and debugging tools at Microsoft, so that, these checks can be triggered automatically before the on-call engineers are engaged.
    
    \item \textbf{Better tooling} - The knowledge extracted by \softner{} can also be used to improve the existing incident reporting and management tools. \softner{} identified entities can be incorporated into the incident report forms, where some of these entities can even be made mandatory fields. We have already started working on this scenario with feature teams that own incident reporting tools at \CompanyX{}.
    
    \item \textbf{Type-aware models} - The multi-task deep learning architecture used in \softner{} uses both the semantic and the data-type context for entity extraction. As per our knowledge, this is the first usage of a multi-task and type-aware architecture in the software engineering domain. Given that software code and programs are typed, this model can potentially be used in other applications like code summarization.

    \item \textbf{Predictive tasks} - In this work, we have shown that the knowledge extracted by \softner{} can be used to build more accurate machine learning models for incident triaging. Similarly, we can build models to automate other tasks such as severity prediction, root causing, etc.
    
    \item \textbf{Bug reports} - Even though this work is motivated by the various problems associated with incident management, the challenges with lack of structure applies to bug reports as well. We plan to evaluate \softner{} on bug reports at \CompanyX{} and, also, on the publicly available bug report data sets.
\end{enumerate}


\section{Related Work}
\label{sec:related-work}

SoftNER is inspired from prior work in two main areas: software engineering and information retrieval. In this section, we discuss related work in these areas.

\textbf{Incident management}: Recent work on incident management has been focused on the problem of triaging incidents to correct teams. As per the empirical study by Chen et al. \cite{EmpiricalIcMICSE2019}, mistriaging of incidents happen quite frequently and can lead up to a delay of over 10X in triaging time, apart from the lost revenue and customer impact. To solve this problem, they have also proposed DeepCT \cite{ContinuousTriageASE2019}, a deep learning based method for routing of incidents to the correct teams. There has also been a significant amount of work done on diagnosing and root causing of service incidents. DeCaf \cite{bansal2019decaf} uses random forest models for automatically detecting and categorizing performance issues in large-scale cloud services. Systems such as AirAlert \cite{chen2019outage} have been built using machine learning to predict critical service incidents, called outages, in large scale services. In our work, we focus on structured knowledge extraction and how it can be used as strong features to build significantly more accurate models for these incident management tasks, as in Section 5.4.

\textbf{Bug reports}: Significant amount of research has been done on bug reports in the traditional software context. \softner{} is inspired by InfoZilla \cite{bettenburg2008extracting} which leverages heuristics and regular expressions for extracting four elements from Eclipse bug reports: patches, stack traces, code, and lists. Unlike InfoZilla, we build a completely unsupervised deep learning based framework that enables \softner{} to extract hundreds of entities without requiring any prior knowledge about them. Our work also targets incidents, which are more complex than bugs because of numerous layers of dependencies, and also, the real-time mitigation requirements. Similar to incidents, existing work on bug reports \cite{anvik2006should}, \cite{tian2016learning} have largely used the unstructured attributes like bug description as it is. Though we have focused on incidents, \softner{} can be applied to bug reports for extracting structured information and building models for tasks like bug triaging, classification, etc.

\textbf{Information retrieval}: Knowledge and entity extraction has been studied in depth by the information retrieval community. Search engines like Google and Bing rely heavily on entity knowledge bases for tasks like intent understanding \cite{pantel2012mining} and query reformulation \cite{xu2008entity}. Supervised methods \cite{florian2003named}, \cite{mccallum2003early} require a large amount of training data which can be cost prohibitive to collect. Hence, search engines commonly use semi-supervised methods which leverage a small seed set to bootstrap the entity extraction process. For instance, the expert editors would seed the entity list for a particular entity type, let's say fruits with some initial values such as \{apple, mango, orange\}. In this work, our goal was to build a fully unsupervised system where we don't need any pre-existing list of entity types or seed values. This is primarily because every service and organization is unique and manually bootstrapping \softner{} would be very laborious. Additionally, incident reports, unlike web data, contain not just natural language tokens but also other entities such as GUIDs and IP Addresses. Hence, SoftNER leverages a novel data-type aware deep learning model for knowledge extraction.
\section{Conclusion}
\label{sec:conclusion}

Incident management is a key part of building and operating large-scale cloud services. In this paper, we propose \softner{}, an unsupervised framework for mining knowledge graphs from incident reports that incorporates a novel multi-task BiLSTM-CRF model for software named-entity recognition. We have evaluated \softner{} on the incident data from \CompanyX{}, a major cloud service provider. Our evaluation shows that even though \softner{} is fully unsupervised, it has a high precision of 0.96 (at rank 50) for learning entity types from the unstructured incident data. Further, our multi-task model architecture outperforms existing state-of-the-art models in entity extraction. Lastly, our novel approach for mining entity relations has a high accuracy of 0.9. We have deployed \softner{} at \CompanyX{}, where it has been used for knowledge extraction from incidents for over 6 months. Lastly, we show that the extracted knowledge can be used for building significantly more accurate models for critical incident management tasks like triaging.


\bibliographystyle{IEEEtran}
\bibliography{references}

\end{document}